\documentclass[12pt, a4paper]{article}
\pdfoutput=1
\usepackage{graphicx}
\usepackage{amssymb}
\usepackage{amsmath}
\usepackage{bm}
\usepackage{color}
\usepackage{cite}
\usepackage{epstopdf}            
\usepackage{epsfig}
            
\newcommand{\beq}{\begin{equation}}
\newcommand{\eeq}{\end{equation}}
\newcommand{\ov}{\overline}

\usepackage[colorlinks=true, linkcolor=black, citecolor=black,
urlcolor=black]{hyperref}

\begin{document}

\begin{titlepage}

%\begin{flushright}
%\end{flushright}

\vskip 2cm
\begin{center}

{\Large
{\bf 
LHCb anomaly and $B$ physics \\ in flavored $Z^{\prime}$ models with flavored Higgs doublets
}
}

\vskip 2cm

P. Ko$^{1}$,
Yuji Omura$^{2}$,
Yoshihiro Shigekami$^3$,
and
Chaehyun Yu$^{4}$

\vskip 0.5cm

{\it $^1$School of Physics, KIAS, Seoul 02455, Korea
}\\[3pt]
{\it $^2$
Kobayashi-Maskawa Institute for the Origin of Particles and the
Universe, \\ Nagoya University, Nagoya 464-8602, Japan}\\[3pt]
{\it $^3$
Department of Physics, Nagoya University, Nagoya 464-8602, Japan}\\[3pt]
{\it $^{4}$
Department of Physics, Korea University, Seoul 02841, Korea}\\ [3pt]

\vskip 0.5cm

\begin{abstract}
We study an extended Standard Model with a gauged U(1)$^{\prime}$ flavor symmetry, motivated not only by the fermion mass hierarchy but also by the excesses in $B \to K^{(*)} l  l$ reported by the LHCb collaborations. The U(1)$^{\prime}$ charges are assigned to quarks and leptons in a flavor-dependent manner, and flavored Higgs doublets are also introduced in order to detail the Yukawa couplings at the renormalizable level. Then, the fermion mass hierarchy is realized by the vacuum alignment of the Higgs doublets. In this model, flavor-changing currents involving the gauge boson of U(1)$^{\prime}$ and the scalars generated by the Higgs doublets are predicted and the observables in the $B \to K^{(*)} l  l$ process possibly deviate from the Standard Model predictions. We study the possibility that these new flavor-changing interactions can explain the excesses in the $B \to K^{(*)} l  l$ process, and we derive some predictions for the other flavor-violating processes based on the analysis. We specifically investigate the $\Delta F=2$ processes and the other $B$ decays: e.g., $B \to X_s  \gamma$ and $B \to D^{(*)} \tau \nu$, where the deviations are reported by the Belle and $BABAR$ collaborations.
\end{abstract}

\end{center}
\end{titlepage}

%%%%%%%%%%%%%%%%%%%%%%%%%%%%%%%%%%
\section{INTRODUCTION}
\label{sec;intro}
%%%%%%%%%%%%%%%%%%%%%%%%%%%%%%%%%%
The fermion mass hierarchy and the flavor mixing in the Standard Model (SM) are mysteries in elementary particle physics, the origin of which one would like to understand. Top quarks are much heavier than other fermions, and bottom quarks and $\tau$ leptons are also relatively heavy. On the other hand, electrons and up and down quarks are much lighter than the other particles. Three active neutrinos are much lighter than even electrons. Not only the mass spectra but also the flavor mixings show interesting patterns. Flavor mixing in the quark sector shows hierarchical structures, whereas those in the leptonic sector show large mixings.

This hierarchical structure in the mass spectra may be a hint of new physics beyond the SM. One well-known solution to explain the mass hierarchy is the Froggatt-Nielsen (FN) mechanism~\cite{FN}. In this mechanism, flavor-dependent U(1)$^{\prime}$ symmetry is assigned to the SM fermions and the fermion mass hierarchy is realized by flavor-dependent suppressions generated by the flavor symmetry. The suppressions come from nonrenormalizable higher-dimensional operators, and the charge assignment of U(1)$^{\prime}$ is nontrivial. This mechanism is, however, known to explain the hierarchy and the flavor mixings well~\cite{FN}.

Inspired by the FN mechanism, we construct a model with U(1)$^{\prime}$ flavor gauge symmetry. In our model, we also introduce flavored Higgs doublet fields charged under the U(1)$^{\prime}$, and we then detail the Yukawa couplings to generate the quark and the lepton mass matrices at the renormalizable level. Then we propose that the vacuum alignment of the Higgs doublets is the origin of the fermion mass hierarchy. Note that we can derive a setup similar to the FN mechanism, integrating out the extra Higgs fields, so our setup proposes the origin of the higher-dimensional operators in the FN mechanism.

On the other hand, several excesses have been reported in the $B$ decays by the LHCb Collaboration. One is the lepton universality in $B \to K l l$ ($l=e, \mu$)~\cite{Aaij:2014ora}, while another is the angular distribution of $B \to K^* \mu \mu$~\cite{Aaij:2013qta,Aaij:2015oid}. The global fitting analysis on the relevant Wilson coefficients has been done, including the $B \to X_s \gamma$ process. Interestingly, the authors in Refs.~\cite{Descotes-Genon:2013wba,Hiller:2014yaa,Altmannshofer:2015sma,Descotes-Genon:2015uva,Hurth:2016fbr} suggest the sizable operators
\begin{equation}
  C_{9}(\overline{s_L} \gamma_\mu b_L) (\overline{\mu}\gamma^\mu \mu)+C_{10}(\overline{s_L} \gamma_\mu b_L) (\overline{\mu}\gamma^\mu \gamma_5 \mu)+{\rm H.c.}
\end{equation}
in order to explain the excesses. This may be a hint of the new physics behind the hierarchical structure of the fermion masses.

In this work, we consider a charge assignment that left-handed quarks and $\mu$ leptons are charged under U(1)$^{\prime}$, and we discuss the anomaly at the LHCb. In addition, we can expect that such a large new physics effect contributes to the other observables in flavor physics. We investigate the correlations and predictions in several flavor-violating processes. As the authors of Refs.~\cite{Ko1,Ko2,Ko3} pointed out, there are correlations between the $Z^{\prime}$ interaction from gauged $U(1)^{\prime}$ and the scalar interaction from the Higgs fields, if we consider an explicit model with $U(1)^{\prime}$. A large $C_{9}$ and $C_{10}$ might affect not only $B \to K^{(*)} l l$ but also the other processes through the scalar interaction. Some of the interesting physical observables are $R(D)$ and $R(D^{*})$, which measure the flavor universalities in the $B \to D \tau \nu$ and $B \to D^{*} \tau \nu$ processes, respectively. The experimental results at the $BABAR$ experiment deviate significantly from the SM prediction~\cite{Lees:2012xj,Lees:2013uzd}. The results reported by the Belle Collaboration~\cite{Huschle:2015rga,Abdesselam:2016cgx} are closer to the SM predictions, but we still have great tension between the experimental results and the predictions, so these measurements motivate us to consider new particles that couple to bottom, charm quarks and $\tau$ leptons or new physics in a model-independent way~\cite{Datta:2012qk,He:2012zp,Biancofiore:2013ki,Dorsner:2013tla,Sakaki:2013bfa,Abada:2013aba,Hagiwara:2014tsa,Duraisamy:2014sna,Sakaki:2014sea,Bhattacharya:2014wla,Alonso:2015sja,Greljo:2015mma,Calibbi:2015kma,Freytsis:2015qca,Bauer:2015knc,Hati:2015awg,Barbieri:2015yvd,Becirevic:2016hea,Alonso:2016gym,Dumont:2016xpj,Boucenna:2016wpr,Boubaa:2016mgn,Das:2016vkr,Li:2016vvp,Feruglio:2016gvd,Alok:2016qyh,Boucenna:2016qad,Deshpand:2016cpw,Becirevic:2016yqi,Sahoo:2016pet,Hiller:2016kry,Bhattacharya:2016mcc,Ligeti:2016npd,Bardhan:2016uhr,Dutta:2016eml,Bhattacharya:2016zcw,Barbieri:2016las,Hue:2016nya,Ivanov:2017mrj,Dutta:2017xmj,Alonso:2017ktd,Cvetic:2017gkt,Tran:2017udy}. One good candidate is a charged Higgs field, which has been widely discussed~\cite{Crivellin:2012ye,Ko-chargedHiggs,Kim:2015zla,Fajfer:2012jt,Celis:2012dk,Tanaka:2012nw,Crivellin:2013wna,Crivellin:2015hha,Enomoto:2015wbn,Cline:2015lqp,Wang:2016ggf,Celis:2016azn} and is realized in our model as well. We study the compatibility between the excesses of $B \to K^{(*)} l l$ and $B \to D^{(*)} \tau \nu$, together with their consistency with $B \to X_s \gamma$.

This paper is organized as follows. In Sec.~\ref{sec;setup}, we introduce our model with the gauged U(1)$^{\prime}$ flavor symmetry, and we present the $Z^{\prime}$ and scalar couplings with the SM fermions. Then, in Sec.~\ref{sec;flavor} we study flavor physics: $B \to K^{(*)} l l$, $B \to D^{(*)} \tau \nu$, $B \to X_s \gamma$, and so on. In Sec.~\ref{sec;DM}, we introduce extra fields to make the flavor gauge symmetry anomaly-free, and propose dark matter candidates. Section~\ref{sec;summary} is devoted to a summary of our results.

%%%%%%%%%%%%%%%%%%%%%%%%%%%%%%%%%%%%%%%%%%%%%%%%%%%%%%%
\section{FLAVORED $Z^{\prime}$ MODEL}
\label{sec;setup}
%%%%%%%%%%%%%%%%%%%%%%%%%%%%%%%%%%%%%%%%%%%%%%%%%%%%%%%
In this section, we introduce our model with a gauged U(1)$^{\prime}$ flavor symmetry under which the SM fermions are charged. The U(1)$^{\prime}$ charges to the SM fermions are summarized in Table~\ref{table1} and are chosen in a manner such that we can realize the fermion mass hierarchy and the sizable $C_{9, 10}$ for the LHCb anomalies. In principle, there could be several possible charge assignments. The choice in Table~\ref{table1} is motivated by the following points:
\begin{itemize}
\item[(a)] The charges of the right-handed down-type quarks are universal, and $C_{9}$ and $C_{10}$ are generated by the flavor-dependent U(1)$^{\prime}$ charges of left-handed quarks.
\item[(b)] The lepton flavor-violating processes involving electron, such as $\mu \to 3\, e$ and $\mu \rightarrow e  \gamma$, are highly suppressed.
\end{itemize}

\begin{table}[t]
  \begin{center}
    \begin{tabular}{cccccc}
      \hline
      \hline
      Fields & ~~spin~~ & ~~$\text{SU}(3)_c$~~ & ~~$\text{SU}(2)_L$~~  & ~~$\text{U}(1)_Y$~~ & ~~U(1)$^{\prime}$~~ \\ \hline  
      $\Hat Q^{a}_L$ & 1/2 & ${\bf 3}$ & ${\bf 2}$ & $1/6$  & $0$       \\  
      $\Hat Q^{3}_L$ & 1/2 & ${\bf 3}$ & ${\bf 2}$ & $1/6$  & $1$       \\ \hline 
      $\Hat u^{a}_R$ & 1/2 & ${\bf 3}$ & ${\bf 1}$ & $2/3$  & $q_{a}$    \\ 
      $\Hat u^{3}_R$ & 1/2 & ${\bf 3}$ & ${\bf 1}$ & $2/3$  & $1+ q_{3}$ \\ \hline
      $\Hat d^{i}_R$ & 1/2 & ${\bf 3}$ & ${\bf 1}$ & $-1/3$ & $-q_1$     \\ \hline
      $\Hat L^{1}$   & 1/2 & ${\bf 1}$ & ${\bf 2}$ & $-1/2$ & $0$       \\
      $\Hat L^{A}$   & 1/2 & ${\bf 1}$ & ${\bf 2}$ & $-1/2$ & $q_e$     \\ \hline 
      $\Hat e^{1}_R$ & 1/2 & ${\bf 1}$ & ${\bf 1}$ & $-1$   & $-q_1$     \\
      $\Hat e^{A}_R$ & 1/2 & ${\bf 1}$ & ${\bf 1}$ & $-1$   & $q_e-q_2$  \\ \hline 
      $ H^i$        & 0   & ${\bf1}$  & ${\bf 2}$ & $1/2$  & $q_i$      \\
      $ \Phi$       & 0   & ${\bf1}$  & ${\bf 1}$ & $0$    & $q_\Phi$    \\ \hline \hline
    \end{tabular}
  \end{center}
  \caption{The charge assignment of the extra U(1)$^{\prime}$ symmetry. $a$, $A$ and $i$ denote the flavor: $a=1, \, 2$, $A=2, \, 3$ and $i=1,\, 2,\, 3$. $q_2$ is defined as $q_2=q_1+1$. The bold entries ``${\bf 3}$" (``${\bf 2}$") show the fundamental representation of SU(3) (SU(2)) and ``${\bf 1}$" shows singlet under SU(3) or SU(2).}
  \label{table1}
\end{table}

Given the charge assignments in Table~\ref{table1}, we can now detail the Yukawa couplings for the quarks and leptons,
\begin{eqnarray}
  V_{{\rm Y}}&=&y^u_{1a} \overline{\Hat Q^1_L} \widetilde{H^a} \Hat u^a_R + y^u_{2a} \overline{\Hat Q^2_L} \widetilde{H^a} \Hat u^a_R +y^u_{33} \overline{\Hat Q^3_L} \widetilde{H^3} \Hat u^3_R +y^u_{32} \overline{\Hat Q^3_L} \widetilde{H^1} \Hat u^2_R \nonumber \\
  &+&y^d_{ai} \overline{\Hat Q^a_L} H^1 \Hat d^i_R + y^d_{3i} \overline{\Hat Q^3_L} H^2 \Hat d^i_R \nonumber \\
  &+&y^e_{11} \overline{\Hat L^1} H^1 \Hat e^1_R + y^e_{AB} \overline{\Hat L^A} H^2 \Hat e^B_R + {\rm H.c.},
  \label{Yukawa}
\end{eqnarray}
where $a$ and $b$ ($A$ and $B$) are the flavor indexes: $a,\,b=1,\,2$ ($A,\,B=2,\,3$). Depending on the actual values of the U(1)$^{\prime}$ charges, $q_{i}$ and $q_e$, there could be additional Yukawa couplings allowed by the full gauge symmetry. We shall assume that the other Yukawa terms are forbidden by the gauge symmetry, adopting appropriate charge assignments for $q_3$ and $q_e$. Note that $\Hat Q^i_{L}=(\Hat u^i_{L}, \, \Hat d^i_{L})^T$, $\Hat u^i_{R}$, $\Hat d^i_{R}$ and $\Hat e^i_{R}$ are the left-handed quarks, right-handed up-type quarks, right-handed down-type quarks, and right-handed leptons in the flavor basis, respectively. The fields in the mass basis are described by $u^i$, $d^i$ and $e^i$ and correspond to the quarks and leptons as $(u^1, \, u^2, \, u^3)=(u, \, c, \, t)$, $(d^1, \, d^2, \, d^3)=(d, \, s, \, b)$, and $(e^1, \, e^2, \, e^3)=(e, \, \mu, \, \tau)$, respectively.

The mass eigenstates can be defined after the electroweak (EW) and U(1)$^{\prime}$ symmetry breaking. As shown in Table~\ref{table1}, ``three" flavored Higgs doublets, represented as $H^i$, and one U(1)$^{\prime}$-charged singlet scalar ($\Phi$) are introduced, and they break the gauge symmetries by developing nonvanishing vacuum expectation values (VEVs). Only Higgs doublets can spontaneously break both EW and U(1)$^{\prime}$ symmetry, but $\Phi$ is also required to avoid the strong bounds from the constraints on the electroweak precision observables (EWPOs) and the heavy resonance search.

According to the charge assignments in Table~\ref{table1}, the renormalizable scalar potential invariant under the assumed gauge symmetries can be written
\begin{eqnarray}
  V_H &=& m^2_{H_i} |H_i|^2 +m^2_{\Phi} |\Phi|^2 + \lambda^{ij}_H |H_i|^2 |H_j|^2+ \lambda^{i}_{H \Phi} |H_i|^2 |\Phi|^2 + \lambda_{ \Phi} |\Phi|^4 \nonumber \\ 
  &-& A_1 H^\dagger_1H_2 \left ( \Phi \right )^{\frac{q_1-q_2}{q_\Phi}}-A_2 H^\dagger_2 H_3 \left ( \Phi \right )^{\frac{q_2-q_3}{q_\Phi}}-A_3 H^\dagger_1 H_3 \left ( \Phi \right )^{\frac{q_1-q_3}{q_\Phi}} + {\rm H.c.},
  \label{eq;spote}
\end{eqnarray}
where $q_\Phi$ is the charge of $\Phi$. In order to realize the fermion mass hierarchy through the Higgs VEVs, we require that the size of each Higgs VEV satisfies the following relation:
\begin{equation}
  \langle H_1 \rangle \ll  \langle H_2 \rangle \ll \langle H_3 \rangle.
  \label{VEV}
\end{equation}
Let us define $q_\Phi$ as $-1$ and assume that $m_{H_1}$ is much heavier than the EW scale. Then $H_1$ can be integrated out and the effective lagrangian is the two Higgs doublet model (2HDM) with $H_{2,3}$ and $\Phi$. The Yukawa couplings can be then described by replacing $H_1$ with the higher dimensional term involving $\Phi$ as follows:
\beq
H_1 \to \frac{A_1}{m^2_{H_1}} \Phi H_2.
\eeq
Note that $H_{2}$ and $H_3$ as well as $\Phi$ will develop nonvanishing VEVs. Then, we define the vacuum alignment of the neutral components as
\beq
\label{eq;vevs}
\langle H_2 \rangle =\frac{v}{\sqrt 2} \cos \beta, \, \langle H_3 \rangle =\frac{v}{\sqrt 2} \sin \beta, \, \langle \Phi \rangle =\frac{v_{\Phi}}{\sqrt 2},
\eeq
and discuss the phenomenology, depending on the vacuum alignment. Note that the explanation of the fermion mass hierarchy through the VEVs requires large $\tan \beta$.

This explanation of the mass hierarchy relies on the scalar potential, specifically, the masses squared of the Higgs fields. In order to realize a small $\langle H_1 \rangle$, $m^2_{H_1}$ needs to be large compared to $A_1 v_{\Phi}$. On the other hand, we would need a large $m^2_{H_2}$ and $m^2_{H_3}$ as well if $v_{\Phi}$ is much larger than the EW scale. Through the $A_2$ term, $v_{\Phi}$ contributes to the stationary conditions for $\langle H_2 \rangle$ and $\langle H_3 \rangle$, so that we need fine-tuning between $m^2_{H_2}$ ($m^2_{H_3}$) and $A_2 v_{\Phi} \tan \beta$ ($A_2 v_{\Phi}/ \tan \beta$) to realize the EW scale. A detailed analysis of the scalar potential has been done in Ref.~\cite{Ko:2013zsa}. Note that dimensionless couplings, such as $\lambda^{ij}_H$ in Eq.~(\ref{eq;spote}), could also play a role in the realization of vacuum alignment. The dimensionless couplings are, however, constrained by the vacuum stability and could easily modify the vacuum alignment according to the radiative corrections if they are ${\cal O} (1)$. Thus, we simply assume that the Higgs masses squared give the mass hierarchy and that $\lambda^{ij}_H$ is approximate to $\delta^{ij} \lambda^i_H$, where $\lambda^i_H$ is not large.

Let us briefly comment on the origin of such specific mass terms of Higgs fields as well, although it is beyond our scope. As mentioned above, the vacuum alignment in Eq.~(\ref{eq;vevs}) is given by the masses squared in the scalar potential, so the Higgs masses in Eq.~(\ref{eq;spote}) need to be hierarchical. One way to explain the origin of the hierarchical Higgs mass terms is to consider the supersymmetric (SUSY) extension of this model with gauged flavor symmetry. In such a SUSY model, $m^2_{H_i}$ corresponds to the soft SUSY breaking parameters, and they are expected to be generated dynamically. In fact, the nonvanishing D-term of the gauged U(1)$^{\prime}$ flavor symmetry can lead to the hierarchical structure of $m^2_{H_i}$, according to the flavor-dependent U(1)$^{\prime}$ charges. In addition, the SUSY extension makes our model stable against radiative corrections: there is no quadratic divergence in the Higgs masses. The fine-tuning to realize the EW scale, however, cannot be avoided---even in the SUSY model---if the SUSY breaking scale is much higher than the EW scale. Depending on the size of $v_{\Phi}$, which is related to the SUSY breaking scale, relaxing the fine-tuning is one of the big issues in our model.

In our study, we do not take into account the origin of $m^2_{H_i}$, so the U(1)$^{\prime}$ charges, $q_i$, $q_e$ and $q_\Phi$, are free parameters in our setup. In order to evade the strong bound on $Z^{\prime}$ mass and U(1)$^{\prime}$ coupling from the Drell-Yan process and allow $A_{1,2}$ terms at the renormalizable level, we define them as $(q_1, \, q_2, \, q_3, \, q_\Phi)=(0,\,1, \, 3, \, -1)$. Note that $q_3$ forbids the $A_3$ term at the renormalizable level. We will choose $q_e$, which can enhance $B \to K^{(*)} l l$.

\subsection{The fermion mass matrices}
After the EW and U(1)$^{\prime}$ symmetry breaking, the mass matrices for the quarks and charged leptons are generated as follows:
\begin{equation}
  \frac{v}{\sqrt 2}Y^u_{ij}  \overline{\Hat u^i_L}  \Hat u^j_R+ \frac{v}{\sqrt 2} Y^d_{ij} \overline{\Hat d^i_L}  \Hat d^j_R+ \frac{v}{\sqrt 2} Y^e_{ij} \overline{\Hat e_L^i}  \Hat e^j_R,
\end{equation}
where each matrix, $Y^{u,d,e}_{ij}$, is defined as
\beq
\left (Y^u_{ij} \right) =  \begin{pmatrix} y^u_{11} \epsilon & y^u_{12} & 0 \\ y^u_{21} \epsilon & y^u_{22} & 0 \\ 0 & y^u_{32} \epsilon & y^u_{33} \\ \end{pmatrix} \begin{pmatrix} \cos \beta && \\ &\cos \beta & \\ && \sin \beta \end{pmatrix}, 
\eeq
\beq
\left (Y^d_{ij} \right) = \cos \beta \begin{pmatrix} \epsilon && \\ & \epsilon & \\ && 1 \end{pmatrix} \begin{pmatrix} y^d_{11} & y^d_{12} & y^d_{13} \\  y^d_{21} & y^d_{22} & y^d_{23} \\  y^d_{31} & y^d_{32} & y^d_{33} \\ \end{pmatrix},
\eeq
and
\beq
\left (Y^e_{ij} \right) = \cos \beta \begin{pmatrix} \epsilon && \\ & 1 & \\ && 1 \end{pmatrix} \begin{pmatrix} y^e_{11} & 0 & 0 \\ 0 & y^e_{22} & y^e_{23} \\ 0 &y^e_{32} & y^e_{33} \\ \end{pmatrix}.
\eeq
$\epsilon$ comes from the contributions of $\langle H_1 \rangle$, and is given by
\beq
\epsilon = \frac{A_1}{m^2_{H_1}} \langle \Phi \rangle. 
\eeq
Using the diagonalizing unitary matrices, the mass matrices are given by
\beq
\label{eq;diagonalization}
\frac{v}{\sqrt{2}} Y^I=(U^{I}_{L})^\dagger {\rm diag} (m^I_1, \, m^I_2, \, m^I_3) U^I_R ~~ (I=u,\,d, \,e).
\eeq

Assuming the hierarchical VEV alignment in Eq.~(\ref{VEV}), the fermion mass hierarchy can be obtained. For instance, the ratios of the up-type quark masses are approximately evaluated from $Y^u$:
\beq
m_c/m_t = {\cal O} (y^u_{22}/y^u_{33} \tan \beta), \, m_u/m_c = {\cal O} (\epsilon \, y^u_{11}/y^u_{22}).
\eeq
A large tan $\beta$ and a small $\epsilon$, corresponding to a small $\langle H_1 \rangle$, can realize the mass hierarchy in the up-type quark sector. Moreover, a small $\epsilon$ can explain the mass hierarchy in the down-type quark and lepton sector:
\beq
m_s/m_b = {\cal O} (\epsilon), \, m_e/m_\mu = {\cal O} (\epsilon).
\eeq
We still need some tuning of the parameters for the hierarchies between strange and down quarks ($\tau$ and $\mu$ leptons) and, especially, the (1, 2) elements of the Cabibbo-Kobayashi-Maskawa (CKM) matrix, but this charge assignment and setup can fit the realistic mass matrices to a good approximation, and it can evade the strong constraints from flavor physics suppressing the tree-level flavor-changing neutral currents (FCNCs), as we will see below.

We can also estimate $U^I_{L}$ and $U^I_{R}$ in Eq. (\ref{eq;diagonalization}). They are relevant to the FCNC interactions of the $Z^{\prime}$ and scalar bosons. Based on the above discussion, we can estimate the size of each element of the unitary matrices according to the hierarchical structures. Those elements that are important in our study on flavor physics are given by 
\beq
\label{eq;UuL}
|(U^d_L)_{33}| \simeq 1 , \, |(U^d_L)_{23}| = {\cal O}(\epsilon) , \, |(U^d_L)_{13}| = {\cal O}(\epsilon)
\eeq
and
\beq
\label{eq;UuR}
|(U^u_R)_{33}| \simeq 1, \,  |(U^u_R)_{23}| = {\cal O}(\epsilon) , \,  |(U^u_R)_{23}| \gg | (U^u_R)_{13}|.
\eeq

\subsection{$Z^{\prime}$ couplings }
Let us discuss the gauged U(1)$^{\prime}$ interaction in this subsection. Based on Table~\ref{table1}, the $Z^{\prime}$ gauge couplings in the flavor basis are given by
\begin{eqnarray}
  {\cal L}_{Z^{\prime}} &=& g^{\prime} \Hat Z^{\prime}_{\mu} \left ( \overline{\Hat Q^3_L} \gamma^\mu \Hat Q^3_L+q_1 \overline{\Hat u^1_R} \gamma^\mu \Hat u^1_R +(1+q_1) \overline{\Hat u^2_R} \gamma^\mu \Hat u^2_R + (1 + q_3) \overline{\Hat u^3_R} \gamma^\mu \Hat u^3_R \right ) \nonumber \\
  &+&g^{\prime} \Hat Z^{\prime}_{\mu} \left (q_e \overline{\Hat L^A} \gamma^\mu \Hat L^A - q_1 \overline{\Hat d^i_R}  \gamma^\mu \Hat d^i_R -q_1 \overline{\Hat e^1_R} \gamma^\mu \Hat e^1_R +(q_e-q_2) \overline{\Hat e^A_R} \gamma^\mu \Hat e^A_R \right ).
  \label{eq;LagZpri}
\end{eqnarray}
After the EW and U(1)$^{\prime}$ symmetry breaking, we obtain the mass eigenstates and the $Z^{\prime}$ couplings to the SM fermions in the mass eigenstates are described as
\begin{eqnarray}
  {\cal L}_{Z^{\prime}} &=& g^{\prime} \Hat Z^{\prime}_{\mu} \left \{ (g^u_L)_{ij} \overline{ u^i_L} \gamma^\mu u^j_L + (g^d_L)_{ij} \overline{d^i_L} \gamma^\mu d^j_L + (g^u_R)_{ij} \overline{u^i_R} \gamma^\mu u^j_R - q_1 \overline{d^i_R} \gamma^\mu d^i_R \right \} \nonumber \\
  &+&g^{\prime} \Hat Z^{\prime}_{\mu} \left \{ q_e \left ( \overline{\mu_L} \gamma^\mu \mu_L +\overline{\tau_L} \gamma^\mu \tau_L \right ) + (g^\nu_L)_{ij} \overline{\nu^i_L} \gamma^\mu \nu^j_L - q_1 \overline{e^1_R} \gamma^\mu e^1_R + (q_e-q_2) \overline{e^A_R} \gamma^\mu e^A_R \right \}. \nonumber \\ \label{eq;18}
\end{eqnarray}
Each coupling in Eq.~(\ref{eq;18}) is defined as
\begin{eqnarray}
  (g^d_L)_{ij} &=& (U^{d}_{L})_{i3} (U^{d}_{L})^*_{j3}, \\
  (g^u_L)_{ij} &=& (U^{u}_{L})_{i3} (U^{u}_{L})^*_{j3} = (V_{\rm CKM})_{ik} (g^d_L)_{kk'} (V_{\rm CKM})^*_{jk'},  \\
  (g^u_R)_{ij} &=& (U^{u}_{R})_{ik} q_k (U^{u}_{R})^*_{jk}, \\
  (g^\nu_L)_{ij} &=& q^k_e \left \{ (U^{\nu}_{L})_{ik} (U^{\nu}_{L})^*_{jk} \right \} = q_e \left \{ \delta_{ij} - (V^\dagger_{\rm PMNS})_{i3} (V^\dagger_{\rm PMNS})^*_{j3} \right \}.
\end{eqnarray}
Note that the Glashow-Iliopoulos-Maiani mechanism does not work here for $Z^{\prime}$ gauge interactions, since U(1)$^{\prime}$ gauge symmetry is flavor-dependent in our model. There will be generic FCNC interactions involving $Z^{\prime}$ in the mass eigenstates of the SM fermions. Based on the estimation of the diagonalizing matrices given in Eqs.~(\ref{eq;UuL}) and (\ref{eq;UuR}), we find that the FCNCs are roughly estimated as 
\begin{eqnarray}
  (g^d_L)_{sb}&=& {\cal O} (\epsilon), \, (g^d_L)_{db}= {\cal O} (\epsilon), \, (g^d_L)_{sd}= {\cal O} (\epsilon^2), \nonumber \\
  (g^u_L)_{ij} & \simeq & (g^d_L)_{ij}, \, (g^u_R)_{ct} =q_3 \times {\cal O} (\epsilon), \, |(g^u_R)_{ct}| \gg |(g^u_R)_{ut}|, \, |(g^u_R)_{uc}|. 
  \label{eq;gdL} 
\end{eqnarray}

In addition, $\Hat Z^{\prime}_{\mu}$ mixes with $\Hat Z_{\mu}$, originated from SU(2)$_L \times$U(1)$_Y$, through the mass mixing generated by nonvanishing VEVs of $H_i$. The mixing is suppressed by $v_\Phi$ and should be tiny in order to evade the strong bound concerned with the $\rho$ parameter. Such a tiny mixing can be achieved by the assumption that $v_\Phi$ dominates the $Z^{\prime}$ mass ($M_{Z^{\prime}}$).

Assuming the U(1)$^{\prime}$ coupling $g^{\prime}$ is comparable with the $Z$ boson coupling, we find that the constraint from the $\rho$ parameter leads to the bound on $\Lambda_{Z^{\prime}} \equiv M_{Z^{\prime}}/g^{\prime}$:
\beq
\label{eq;rho-para}
\Lambda_{Z^{\prime}} \gtrsim 23.3 \, {\rm TeV} \times \left ( \frac{q_3}{3} \right ) \times \left ( \frac{ 10^{-3}}{ \Delta \rho_{\rm max}} \right )^{\frac{1}{2}},
\eeq
when $\tan \beta$ is large. $\Delta \rho_{\rm max}$ is the upper bound on the $\rho$ parameter and roughly estimated as ${\cal O}(10^{-3})$~\cite{PDG}.

In addition, kinetic mixing between U(1)$^{\prime}$ and U(1)$_Y$ is also allowed by the gauge symmetries~\cite{Holdom:1985ag}. This might originate from grand unification or can be generated by the one-loop exchange of extra scalars to the $Z^{\prime}$-$\gamma$ ($Z$) interaction~\cite{Ko:2014uka}. Even if we tune the kinetic mixing to be vanishing at some scale, the renormalization group (RG) correction would induce the finite mixing, which is linear to $g^{\prime}$ and suppressed by the U(1)$_Y$ gauge coupling and the loop factor, at the low scale. The kinetic mixing may cause tension with the EWPOs. The constraint on the kinetic mixing is roughly the same as the mass mixing, so that the kinetic mixing term, $\epsilon_Y F^{\mu \nu}_Y F^{\prime}_{\mu \nu}$, should be tuned to be below $|\epsilon_Y| M_Z^2/M_{Z^{\prime}}^2 \lesssim {\cal O} (10^{-3})$ at the EW scale~\cite{Hook:2010tw}. $M_{Z^{\prime}}^2$ is expected to be, at most, ${\cal O} (10)$ TeV in order to avoid too large of a RG correction.

\subsection{Yukawa couplings}
After the EW and U(1)$^{\prime}$ symmetry breaking, a number of scalar bosons appear as physical modes from the Higgs fields. In the limit where the scalars of $\Phi$ are much heavier than the others, we find two $CP$-even scalars, $h$ and $H$, a $CP$-odd scalar, $A$, and a charged Higgs fields, $H^{\pm}$. There are also extra $CP$-even and $CP$-odd scalars from the fields charged under U(1)$^{\prime}$, such as the $\Phi$, $\Phi_{l}$, and $\Phi_{r}$ introduced in Tables~\ref{table1} and \ref{table4}. Assuming that the mixing between the scalar fields from the Higgs fields and the SM-singlet fields is not significantly large, the Yukawa couplings of these scalar bosons with the SM fermions in the mass basis are given by
\begin{eqnarray}
  -{\cal L}_{Y} &=& (Y^{u}_{S})_{ij} S \overline{ u^i_L} u^j_R + (Y^{d}_{S})_{ij} h  \overline{ d^i_L} d^j_R + (Y^{e}_{S})_{ij} H \overline{ e^i_L} e^j_R \nonumber \\
  &+& (Y^{u}_{\pm})_{ij} H^- \overline{ d^i_L} u^j_R + (Y^{d}_{\pm})_{ij} H^+ \overline{ u^i_L} d^j_R + (Y^{e}_{\pm})_{ij} H^+ \overline{ \nu^i_L} e^j_R+ {\rm H.c.},   
\end{eqnarray}
where $S$ denotes three neutral scalar fields: $S=h, \, H, \, A$, and $H^\pm$ denotes the charged Higgs fields.

In our model, each Yukawa coupling is given as follows:
\begin{eqnarray}
  (Y^{u}_{h})_{ij} &=& \frac{m^i_u \sin (\alpha -\beta )}{v} G_{ij} + \frac{m^i_u \cos (\alpha -\beta )}{v} \delta_{ij}, \\
  (Y^{u}_{H})_{ij} &=& \frac{m^i_u \cos (\alpha -\beta )}{v} G_{ij} - \frac{m^i_u \sin (\alpha -\beta )}{v} \delta_{ij}, \\
  (Y^{u}_{A})_{ij} &=& -i \frac{m^i_u}{v} G_{ij}, \\
  (Y^{u}_{\pm})_{ij} &=& - \frac{m^k_u \sqrt{2}}{v} V^*_{ki} G_{kj},
\end{eqnarray}
and
\begin{eqnarray}
  (Y^{d}_{h})_{ij} &=& - \delta_{ij} \, \frac{m^i_d}{v} \frac{\cos \alpha}{\cos \beta}, \\
  (Y^{d}_{H})_{ij} &=&\delta_{ij} \, \frac{m^i_d}{v} \frac{\sin \alpha}{\cos \beta}, \\
  (Y^{d}_{A})_{ij} &=& -i \delta_{ij} \, \frac{m^i_d}{v} \tan \beta, \\
  (Y^{d}_{\pm})_{ij} &=& - V_{ij} \frac{ m^j_d \sqrt 2}{v} \tan \beta.
\end{eqnarray}
$(Y^{e}_{S})_{ij}$ and $(Y^{e}_{\pm})_{ij}$ are given by replacing $m^i_d$ and $V_{ij}$ by $m^i_e$ and $(V_{\rm PMNS})^*_{ji}$ in $(Y^{d}_{S})_{ij}$ and $(Y^{d}_{\pm})_{ij}$, respectively. $G_{ij}$ originates from the flavor gauge symmetry and is described as
\begin{eqnarray}
  \displaystyle
  G_{ij}&=&\left ( U^u_R \begin{pmatrix} - \tan \beta && \\ & -\tan \beta & \\ && \frac{1}{\tan \beta} \end{pmatrix} U^{u \, \dagger}_R \right )_{ij} \nonumber \\
  &=&- \tan \beta \, \delta_{ij}+ \left ( \tan \beta + \frac{1}{\tan \beta} \right ) (G^u_R)_{ij}, 
\end{eqnarray}
where $(G^u_R)_{ij}$ is defined by $(G^u_R)_{ij} \equiv (U^u_R)_{i3} (U^u_R)^*_{j3}$. Because $G_{ij} \propto \delta_{ij}$ is no longer true in the presence of flavor-dependent U(1)$^{\prime}$ gauge interactions, there appear to be nonminimal flavor-violating scalar interactions, which were first noticed in Refs.~\cite{Ko1,Ko2}. These new interactions are absent in the usual 2HDMs with softly broken $Z_2$ symmetries, and they could play an important role in understanding the flavor-nonuniversal phenomena through flavor-violating gauge and scalar interactions~\cite{Ko1,Ko2,Ko3}. Based on the estimation of $(U^u_R)_{i3}$ in Eq.~(\ref{eq;UuR}), we obtain
\beq
\label{GuR}
(G^u_R)_{tt} \simeq 1, \, |(G^u_R)_{tc}| \simeq  {\cal O} (\epsilon), \,   |(G^u_R)_{tc}| \gg |(G^u_R)_{ut}|, \, |(G^u_R)_{uc}|.
\eeq
Then we find that $(Y^u_{S})_{tc}$ and $(Y^u_{\pm})_{bc}$ are relatively larger and the elements other than $(Y^u_{S})_{tt}$ are highly suppressed. This is very interesting because the Belle and $BABAR$ collaborations reported some excess in $B \to D^{(*)} \tau \nu$~\cite{Lees:2013uzd,Huschle:2015rga,Abdesselam:2016cgx}. In Sec.~\ref{sec;flavor}, we will study this excess together with relevant observables.

Note that the constraint on the EWPOs as well as the Higgs signals has been investigated at the one-loop level in 2HDM with U(1)$^{\prime}$ gauge symmetry, where the Higgs fields are charged under U(1)$^{\prime}$~\cite{Ko:2013zsa}. The degenerate spectrum of the scalar fields and $\cos (\alpha - \beta) \sim 1$ are also required, in addition to the constraint in Eq.~(\ref{eq;rho-para}). In our analysis of flavor physics, we assume that the scalar fields, except for $h$, are almost degenerate and $\cos (\alpha - \beta)$ is close to unity.

%%%%%%%%%%%%%%%%%%%%%%%%%%%%%%%%%%
\section{FLAVOR PHYSICS INVOLVING $b$ QUARK}
\label{sec;flavor}
%%%%%%%%%%%%%%%%%%%%%%%%%%%%%%%%%%
Based on the setup and interaction Lagrangians derived in the previous section, we shall study the relevant flavor physics:  $B \to K^{(*)} l  l$, the $\Delta F=2$ processes, $B \to X_s \gamma$, and $B \to D^{(*)} \tau \nu$. The input parameters to be used are summarized in Table~\ref{table;input}.
%%%%%%%%Input parameters%%%%%%%%%%%%%%%%
\begin{table}
  \begin{center}
    \begin{tabular}{|c|c||c|c|} \hline
      $\alpha_s(M_Z)$ & $0.1193(16)$~\cite{PDG}&   $\lambda$& 0.22537(61)~\cite{PDG}   \\ 
      $G_F$  & 1.1663787(6)$\times 10^{-5}$ GeV$^{-2}$~\cite{PDG}   &  $A$& $0.814^{+0.023}_{-0.024}$~\cite{PDG}  \\ 
      $m_{b}$&4.18$\pm 0.03$ GeV~\cite{PDG} &  $\overline{\rho}$& 0.117(21)~\cite{PDG} \\ 
      $m_t$& 160$^{+5}_{-4}$ GeV~\cite{PDG} &   $\overline{\eta}$& 0.353(13)~\cite{PDG}  \\ 
      $m_{c}$&1.275$\pm 0.025$ GeV~\cite{PDG}  &  &  \\  \hline
    \end{tabular}
    \caption{The input parameters relevant to our analyses. The CKM matrix, $V$, is written in terms of $\lambda$, $A$, $\overline{\rho}$ and $\overline{\eta}$~\cite{PDG}.}
    \label{table;input}
  \end{center}
\end{table}
%%%%%%%%%%%%%%%%%%%%%%%%%%%%%%%%%%%%%%%

\subsection{$b \to s l l$ and $\Delta F$=2 processes }
First, we consider the $b \to s l l$ ($l=e, \, \mu$) decays. In this model, tree-level $Z^{\prime}$ exchange diagrams contribute to the flavor-violating processes, $b\to s ll$.  In the $\Delta B = 1$ effective Hamiltonian, the relevant tree-level contributions are given by
\begin{equation}
{\cal H}_\textrm{eff} = -g_{\rm SM} \left [ C_9^{l}(\overline{s_L} \gamma_\mu b_L) (\overline{l}\gamma^\mu l ) + C_{10}^{l}(\overline{s_L} \gamma_\mu b_L) (\overline{l}\gamma^\mu \gamma_5 l) + {\rm H.c.} \right ] ,
\end{equation}
where $C_{9}^{l}$ and $C_{10}^{l}$ are given by
\begin{subequations}
  \begin{eqnarray}
    C_{9}^{e} &=& C_{10}^{e} = \frac{g^{\prime \, 2}}{2 g_{\rm SM} M^2_{Z^{\prime}}} (g^d_L)_{sb} \, q_1 \ , \\
    C_{9}^{\mu} &=& C_{9}^{\tau} = -\frac{g^{\prime \, 2}}{2 g_{\rm SM} M^2_{Z^{\prime}}} (g^d_L)_{sb} \, (2 q_e - q_2) \ , \\
    C_{10}^{\mu} &=& C_{10}^{\tau} = \frac{g^{\prime \, 2}}{2 g_{\rm SM} M^2_{Z^{\prime}}} (g^d_L)_{sb} \, q_2  \ ,
  \end{eqnarray}
  \label{c9c10}
\end{subequations}
and the SM contributions are omitted. $g_{\rm SM}$ is the factor from the SM contribution:
\begin{equation}
  g_{\rm SM} =\frac{4 G_F}{\sqrt{2}} V_{tb} V^*_{ts} \frac{e^2}{16 \pi^2} ,
\end{equation}
and it is real to a good approximation. We note that the Wilson coefficients in Eq.~(\ref{c9c10}) are flavor dependent. Since we set $q_1=0$ and $q_2=1$, the process $b\to s ee$ is not affected by the $Z^{\prime}$ exchange at the tree level. The process $b\to s \tau\tau$ will be also affected in our model, but we do not consider the processes because of the lack of experimental data.

Furthermore, the branching ratio for the process $b\to s \nu\nu$ can also deviate from the $Z^{\prime}$ coupling, $(g^\nu_L)_{ij}$. The current experimental bound is a factor about 4 above the SM prediction~\cite{Lees:2013kla,Lutz:2013ftz}. As discussed below, we require that the $Z^{\prime}$ contribution is about 20 \% of the SM prediction from the analysis in the $b\to s ll$ processes. The magnitude of the corresponding Wilson coefficient for $b\to s \nu\nu$ in the SM is about $-6.4$, whose absolute value is not so different from the ones for $b\to s ll$. Then the bound on new physics for the Wilson coefficient in the $b\to s \nu\nu$ decays is much larger than the value in the SM; for example, the limit on the Wilson coefficient by new physics is about $-23$, assuming that the Wilson coefficient is diagonal and identical in neutrino flavor~\cite{Bhattacharya:2016mcc}. Thus, we can conclude that our model is surely safe from the current experimental bound in the $b\to s\nu\nu$ decay.

\begin{figure}[!t]
  \begin{center}
    {\epsfig{figure=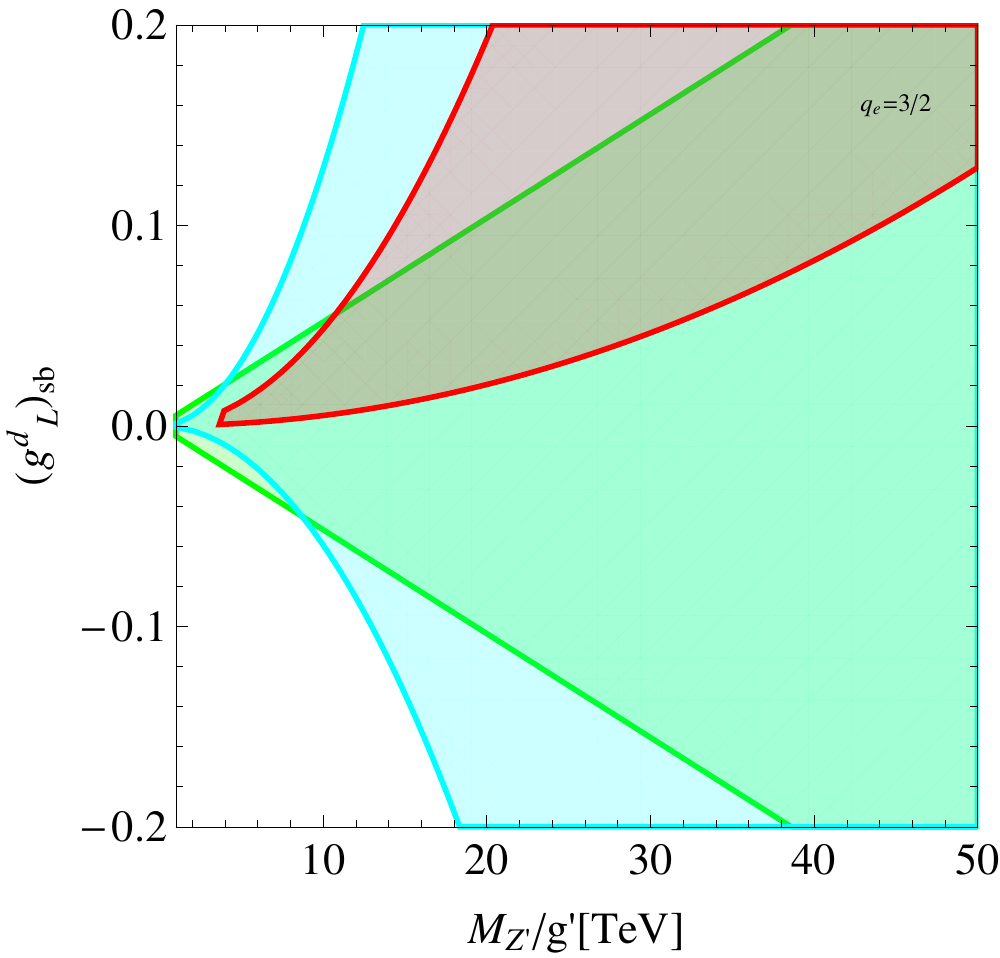,width=0.45\textwidth}}
    {\epsfig{figure=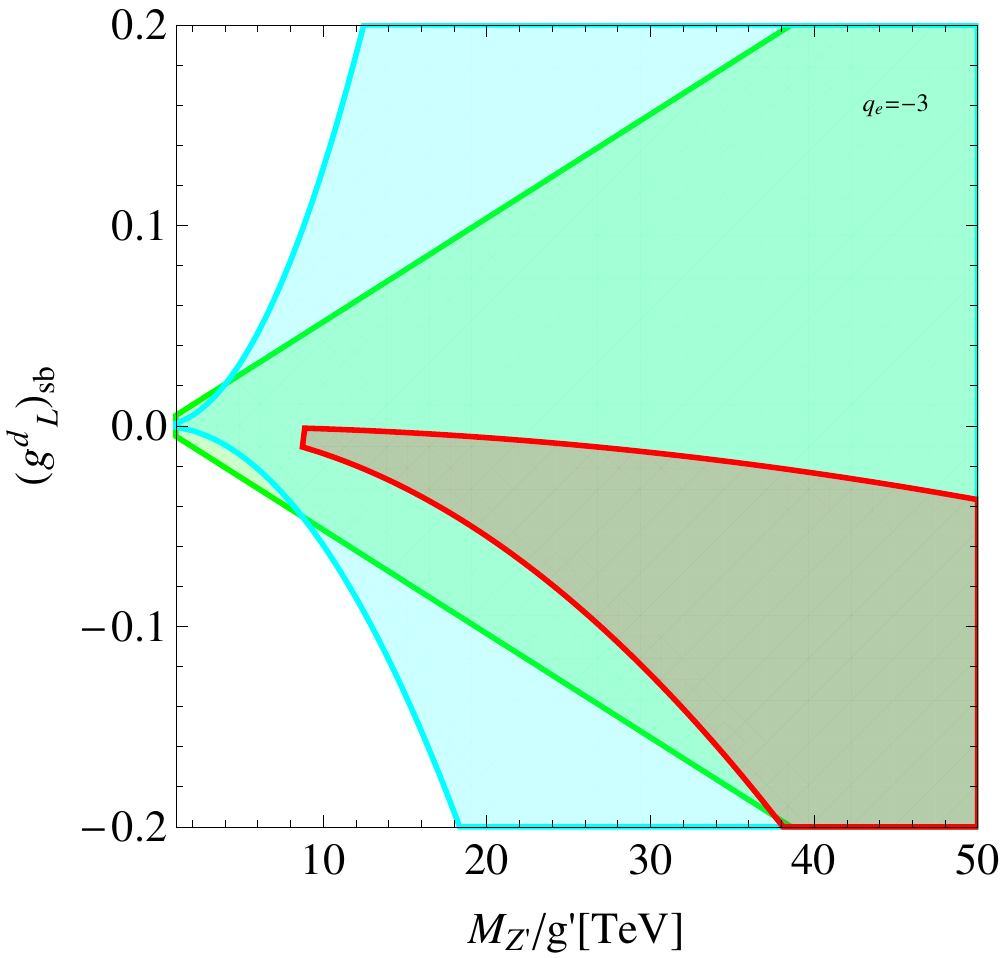,width=0.45\textwidth}}
    \caption{$M_{Z^{\prime}}/g^{\prime}$ vs $(g^d_L)_{sb}$ for $q_e =\frac{3}{2}$ (left panel) and $q_e=-3$ (right panel), respectively. The red and blue regions are allowed by the global fits of $C_9^\mu$ and $C_{10}^\mu$, respectively, within $1\sigma$. The green region is allowed by the $B_s$-$\overline{B}_s$ mixing within $1\,\sigma$.}
    \label{fig;btosll}
  \end{center}
\end{figure}

In the LHCb experiment, a few discrepancies between the experimental results and the SM predictions have been reported in the $b\to s ll$ decays. One of them is the flavor universality in the $B\to K  l  l$ decays~\cite{Aaij:2014ora}, where the discrepancy in the $l=\mu$ and $l=e$ channels is about $2.6\sigma$. Another interesting anomaly which was reported by the LHCb Collaboration is the angular observable $P_5^{\prime}$ in the $B\to K^* \mu^+\mu^-$ with a $3.4\sigma$ deviation~\cite{Aaij:2013qta,Aaij:2015oid}. 

Motivated by these discrepancies, a lot of new physics scenarios have been proposed ---particularly flavor-dependent $Z^{\prime}$ boson scenarios~\cite{Boucenna:2016wpr,Boucenna:2016qad,Chang:2009tx,Chang:2010zy,Chang:2013hba,Crivellin:2015lwa,Altmannshofer:2015mqa,Allanach:2015gkd,GarciaGarcia:2016nvr,Ko:2017quv,Ko:2017yrd}, which is a subject of this work. In order to restrict the Wilson coefficients in Eq. (\ref{c9c10}), we adopt the results of the model-independent analysis in the framework of global fits in the space of Wilson coefficients to available data on the $b\to s l l$ decays, which include  $B\to K l l$, $B\to K^* l l$, $B_s\to \phi l l$, $B_s\to l l$, and/or $b\to s \gamma$ processes~\cite{Descotes-Genon:2013wba,Hiller:2014yaa,Descotes-Genon:2015uva,Altmannshofer:2015sma,Hurth:2016fbr}. We take the result in Ref.~\cite{Hurth:2016fbr} for $C_9^e=C_{10}^e=0$. The ratios of the Wilson coefficients with new physics to those in the SM are allowed in the range of
\begin{subequations}
  \label{globalfits}%
  \begin{eqnarray}
    -0.29\,(-0.34) \le &{C_9^\mu}/{C_9^\textrm{SM}}& \le -0.013\,(0.053), \\
    -0.19\,(-0.29) \le &{C_{10}^\mu}/{C_{10}^\textrm{SM}}& \le 0.088\,(0.15)
  \end{eqnarray}
\end{subequations}
at the $1\sigma$ ($2\sigma$) level, respectively~\cite{Hurth:2016fbr}. In Fig.~\ref{fig;btosll}, we depict the allowed region of $M_{Z^{\prime}}/g^{\prime}$ and $(g_L^d)_{sb}$ for the $q_e=\frac{3}{2}$ case (left panel) and the $q_e=-3$ case (right panel), respectively. The red and blue regions are allowed by the global fits in Eq.~(\ref{globalfits}) for $C_9^{\mu}$ and $C_{10}^\mu$ at the $1\sigma$ level, respectively. As shown in Fig.~\ref{fig;btosll}, the constraint on $C_9^\mu$ is much stronger than the one on $C_{10}^\mu$. Note that there is a lower bound on $M_{Z^{\prime}}/g^{\prime}$ from the $\rho$ parameter, as shown in Eq.~(\ref{eq;rho-para}). For $M_{Z^{\prime}}/g^{\prime} \gtrsim 20$ TeV, we find that the allowed regions require a sizable mixing $|(g^d_L)_{sb}|\gtrsim 0.01$ in both cases.

Next, we consider the $\Delta F=2$ process.  The $Z^{\prime}$-mediated FCNCs are strongly constrained by the $B_s$-$\overline{B}_s$ mixing. The relevant effective Hamiltonian for $\Delta F=2$ with the $Z^{\prime}$ exchange is
\begin{equation}
  {\cal H}^{\Delta F=2}_\textrm{eff}= C^{ij}_1 (\overline{d^i_L} \gamma_\mu d^j_L) (\overline{d^i_L} \gamma_\mu d^j_L), 
  \label{coeffF2}%
\end{equation}
where $C^{ij}_1$ is 
\begin{equation}
  C^{ij}_1 = \frac{g^{\prime \, 2}}{2 M^2_{Z^{\prime}}} (g^d_L)_{ij} (g^d_L)_{ij}.
\end{equation}
The SM contribution $C_1^{ij,\textrm{SM}}$ is omitted again. We note that, in general, the right-handed $Z^{\prime}$ exchange can also generate the $\Delta F=2$ process~\cite{Baek:2008vr,Li:2012xc,Li:2015xna}, but the contribution is suppressed by small mixings in $(g^d_R)_{ij}$ and $(g^u_R)_{ij}$ for $q_i = 0$, as shown in Eqs.~(\ref{eq;18}) and (\ref{eq;gdL}). The heavy Higgs exchange also generates extra contributions, but it could vanish in the SM limit and the assumption of the small scalar mass difference: $\cos (\beta -\alpha)=1$ and $m_H=m_A$. This condition is required by the EWPOs, as mentioned at the end of Sec.~\ref{sec;setup}.

In the SM, the mass difference of the $B_{1,2} \equiv ( B_s \pm \overline{B}_s )/\sqrt{2}$ mesons is obtained as
\begin{equation}
  \Delta m_s = \frac{G_F^2}{6 \pi^2} m_{B_s} f_{B_s}^2 \hat{B}_{B_s} \eta m_W^2 |V_{tb}^* V_{ts}|^2 S_0 (x_t),
\end{equation}
where we ignore the small imaginary part in the CKM matrix elements, and the function $S(x_t)$ with $x_t=m_t^2/m_W^2$ can be found in Ref.~\cite{Buras:1998raa}. The QCD correction factor $\eta=0.551$ and $f_{B_s} \hat{B}_{B_s}^{1/2}= 0.266\pm 0.018$~\cite{Aoki:2016frl}, where $f_{B_s}$ is the decay constant of the $B_s$ meson and $\hat{B}_{B_s}$ is the bag parameter. The averaged value for the measured $\Delta m_s$ is $\Delta m_s = 17.757\pm 0.021$ ps$^{-1}$~\cite{Amhis:2014hma}. We note that the uncertainties in $f_{B_s} \hat{B}_{B_s}^{1/2}$ dominate over those in other parameters as well as in $\Delta m_s$.

In Fig.~\ref{fig;btosll}, the green region is allowed by the $\Delta m_s$ constraint at the $1\sigma$ level. One can find the region which is in agreement with the global fits in the $b\to s  l  l$ processes and $\Delta m_s$ simultaneously for both the $q_e=3/2$ and $q_e=-3$ cases. However, we find that there is no allowed region for small $q_e$'s: for example, $q_e=1/2$ at the $1\,\sigma$ level. For $M_{Z^{\prime}}/g^{\prime} \sim 20$ TeV, the upper bound on $(g_L^d)_{sb}$ is estimated as
\begin{equation}
  0.04\,(0.01) \lesssim |(g_L^d)_{sb}| \lesssim 0.1\,(0.1)
\end{equation}
for both the $q_e=3/2$ ($-3$) cases, respectively.

Since the scalar potential in our model is approximated to be the 2HDM at the EW scale, the charged Higgs boson also contributes to the $B_s$-$\overline{B}_s$ mixing through box diagrams~\cite{Kim:2015zla,Buras:2001mb}. We find that this contribution can be large for a small $\tan \beta$ or a large $(G_R^u)_{tc}$, but it can be negligible for $\tan \beta \gtrsim 5$ and $(G_R^u)_{tc}\lesssim 0.1$.

In addition to the $B_s$-$\overline{B_s}$ mixing, other $\Delta F=2$ processes also require the suppressed off-diagonal elements of $(g^d_L)_{ij}$. Assuming that $(g^d_L)_{ij}$ has the same phase as $(V^*_{ti}V_{tj})^2$, we obtain the strong bounds on the $(d, b)$ and $(s, d)$ elements of $(g^d_L)_{ij}$:
\beq
|(g^d_L)_{db} | \, \lesssim \, 8.54 \times 10^{-4} \, \left (\frac{\Lambda_{Z^{\prime}}}{{\rm TeV}} \right ), \, |(g^d_L)_{sd} | \, \lesssim \, 3.47 \times 10^{-5} \, \left (\frac{\Lambda_{Z^{\prime}}}{{\rm TeV}} \right ).
\eeq
Here, these bounds are given by the requirements that the deviations of $\Delta M_{B_d}$ and $|\epsilon_K|$ from the SM predictions are less than 10 \%. When $\epsilon \sim m_s/m_b $, the small flavor-changing couplings are realized as discussed in Eq.~(\ref{eq;gdL}). The $\Lambda_{Z^{\prime}}= {\cal O} (10)$ TeV scenario can satisfy these strong upper bounds. Note that the constraint from $|\epsilon_K|$ can be drastically relaxed if we simply assume that the imaginary part of $(g^d_L)_{sd}$ is vanishing.
  
Let us give a comment on the contributions of the scalars to $B \to K^* \mu \mu$ in our model. The box diagrams involving the charged Higgs scalar generate the operators, such as
\beq
\Delta {\cal H}^{b - s}_\textrm{eff}= 
C^{\mu}_{LR} (\overline{s_L} \gamma_\mu b_L)(\overline{\mu_R }\gamma^\mu  \mu_R) +C^{\mu}_{RR} (\overline{s_R} \gamma_\mu b_R)(\overline{\mu_R }\gamma^\mu  \mu_R)+ {\rm H.c.} 
\eeq
In addition, there are box diagrams involving both $W$ and charged Higgs bosons. Those operators modify our prediction given by $Z^{\prime}$ exchange. Those Wilson coefficients are, however, suppressed by the CKM matrix and the fermion masses originating from the Yukawa couplings, even though $\tan \beta$ enhances each coefficient. The dominant contribution appears in $C^{\mu}_{LR}$, which is proportional to the top quark mass and the large $G_{tt}$. The $B \to X_s \gamma$ process, however, constrains $G_{tt}$ and $\tan \beta$. In addition, the Yukawa couplings of charged Higgs bosons with muons are suppressed by the muon mass. Therefore, we would not expect the coefficients to be sizable enough to change our prediction to $B \to K^* \mu   \mu$.

\subsection{ $B \to X_s \gamma$ }
Based on the previous discussion, we study other flavor-violating processes: e.g., $B \to X_s \gamma$ and $B \to D^{(*)} \tau \nu$. Specifically, it is known that the former process strongly constrains the extra Higgs contributions.

The branching ratio of $B \to X_s \gamma$ has been calculated in 2HDMs~\cite{Borzumati:1998tg,Borzumati:1998nx,Hermann:2012fc,Misiak:2015xwa}. The relevant operators are 
\begin{equation}
  {\cal H}^{b \to s\gamma}_\textrm{eff}= -\frac{4 G_F}{\sqrt 2} V^*_{ts}  V_{tb} \left ( C_7 {\cal O}_7+ C_8 {\cal O}_8 \right ),
\end{equation}  
where the operators are defined as
\beq
{\cal O}_7= \frac{e}{16 \pi^2} m_b (\overline{s_L} \sigma^{\mu \nu} b_R) F_{\mu \nu}, \, {\cal O}_8= \frac{g_s}{16 \pi^2} m_b (\overline{s_L} t^a \sigma^{\mu \nu} b_R) G^a_{\mu \nu}.
\eeq
In our model, the one-loop corrections involving a charged Higgs appear in $C_7$ and $C_8$:
\begin{eqnarray}
  \label{c7}%
  C_7 &=& \left ( \frac{m^u_j m^u_k}{m^2_t} \right ) \frac{V_{kb} V^*_{js}}{V_{tb} V^*_{ts}} G^*_{ki} G_{ji} C^{(1)}_7 (x_i)+\left ( \frac{m^u_k}{m_t} \right ) \frac{V_{ib} V^*_{ks}}{V_{tb} V^*_{ts}} G_{ki} \tan \beta  \, C^{(2)}_7 (x_i), \\
  C_8 &=& \left ( \frac{m^u_j m^u_k}{m^2_t} \right ) \frac{V_{kb} V^*_{js}}{V_{tb} V^*_{ts}} G^*_{ki} G_{ji} C^{(1)}_8 (x_i)+\left ( \frac{m^u_k}{m_t} \right ) \frac{V_{ib} V^*_{ks}}{V_{tb} V^*_{ts}} G_{ki} \tan \beta  \, C^{(2)}_8 (x_i). 
\end{eqnarray}
where $x_i=(m^u_i/m_{H_{\pm}})^2$ and the loop functions are given by
\begin{eqnarray}
  C^{(1)}_7(x) &=& \frac{x}{72} \left \{ \frac{-8x^3+3x^2+12x -7 +(18x^2 -12x ) \ln x}{(x-1)^4} \right \}, \\
  C^{(2)}_7(x) &=& \frac{x}{12} \left \{ \frac{-5x^2+8x -3 +(6x -4 ) \ln x}{(x-1)^3} \right \}, \\
  C^{(1)}_8(x) &=& \frac{x}{24} \left \{ \frac{-x^3+6x^2-3x -2-6x  \ln x}{(x-1)^4} \right \}, \\
  C^{(2)}_8(x) &=& \frac{x}{4} \left \{ \frac{-x^2+4x -3 -2 \ln x}{(x-1)^3} \right \}.
\end{eqnarray}
Note that the SM contributions are $C^{{\rm SM}}_7 =3 \, C^{(1)}_7 (m^2_t/M^2_W)$ and $C^{{\rm SM}}_8 =3 \, C^{(1)}_8 (m^2_t/M^2_W)$.

The branching ratio of $B \to X_s \gamma$ has been measured and the result is consistent with the SM prediction. If $(G^u_R)_{tt}$ deviates from $1$, the charged Higgs contributions would change the SM prediction drastically for a large $\tan \beta$ because there is a term linear to $\tan \beta$ in $(Y^u_{\pm})_{ij}$.

In Ref.~\cite{Misiak:2015xwa}, the calculation of the SM prediction for $B \to X_s \gamma$ has been done at the next-to-next-to-leading-order level. Following that result, we obtain our prediction for this process. In order to survey the allowed region in our model, we consider two parameter choices in this section and the next:
\begin{itemize}
\item (A) $((G^u_R)_{tt}, \, (G^u_R)_{tc}, \, (G^u_R)_{cc}, \, (G^u_R)_{uu}) = (1-(G^u_R)_{cc}, \, 0.03, \, 10^{-3}, \, 0)$,
\item (B) $((G^u_R)_{tt}, \, (G^u_R)_{tc}, \, (G^u_R)_{cc}, \, (G^u_R)_{uu}) = (1-(G^u_R)_{cc}, \, -0.3, \, 0.1, \, 0)$. 
\end{itemize}
The parameters in case (A) correspond to $\epsilon \simeq 0.03$, which is predicted by $m_s/m_b$, as discussed in Sec.~\ref{sec;setup}. In case (B), $(G^u_R)_{tc}$ is fixed to be $-0.3$, and the behavior of the branching ratio of $B \to X_s  \gamma$ is completely different from case (A).

\begin{figure}[!t]
\begin{center}
{\epsfig{figure=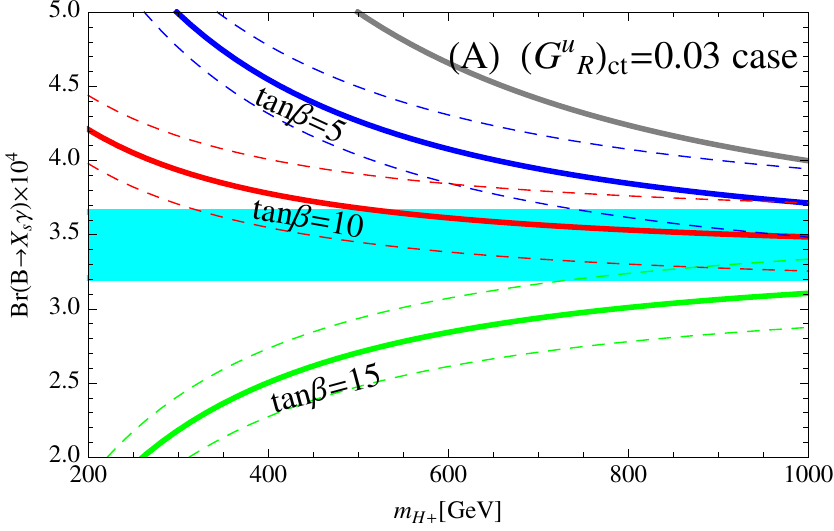,width=0.45\textwidth}}{\epsfig{figure=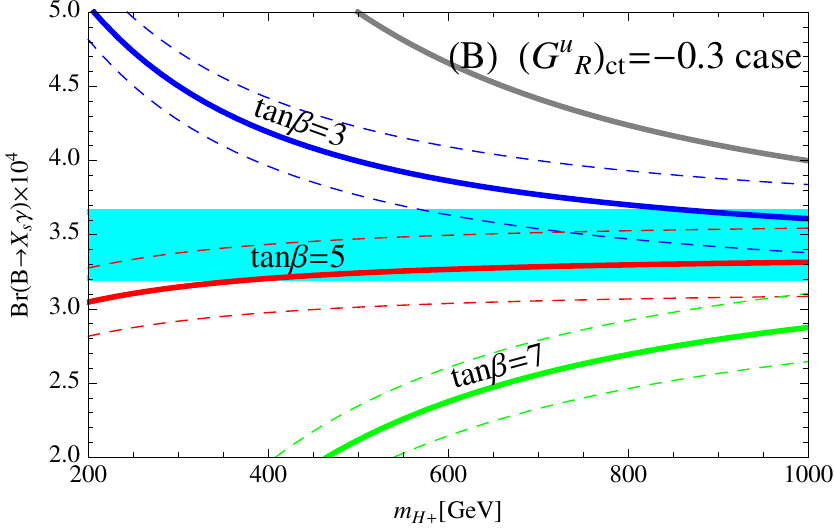,width=0.45\textwidth}} 
\caption{Our predictions of Br($B \to X_s \gamma$) in cases (A) and (B) for some fixed values of $\tan\beta$: $\tan \beta = 5$ (blue line), $10$ (red line), and $15$ (green line) in case (A), and $\tan \beta=3$ (blue line), $5$ (red line), and $7$ (green line) in case (B). The charged Higgs mass is varied within $200\, {\rm GeV} \leq m_{H_{\pm}} \leq 1.0\, {\rm TeV}$.  The cyan region is the $1\sigma$ range of the experimental results~\cite{Amhis:2014hma}. The solid gray line corresponds to the $\tan \beta=50$ case with $(G^u_R)_{ct}=-10^{-3}$.}
\label{fig;BtoXsgamma}
\end{center}
\end{figure}
 
In Fig.~\ref{fig;BtoXsgamma}, we draw the branching ratio for $B\rightarrow X_s \gamma$ in each case. For this process, an important parameter is $(Y^u_{\pm})_{st}$, defined as
\beq
\label{eq;Yst}
(Y^u_{\pm})_{st} \simeq - \frac{m_t \sqrt{2}}{v} V^*_{ts} G_{tt}- \frac{m_c \sqrt{2}}{v} V^*_{cs} G_{ct}.  
\eeq
Note that  the coupling $(Y^u_{\pm})_{st}$ can vanish for different values of $\tan\beta$ from the cancellation of two terms in Eq. (\ref{eq;Yst}),  depending on the sign of $G_{ct}$. 

In case (A), we fix $\tan \beta$ to be $\tan \beta = 5$ (blue line), $10$ (red line), and $15$ (green line), respectively. The predictions are drawn in the left panel of Fig.~\ref{fig;BtoXsgamma}. Dashed lines depict the region including the $\pm 1\sigma$ error of the SM prediction for each $\tan \beta$ case. The cyan band is the $1\sigma$ region of the combined experimental result~\cite{Amhis:2014hma}. In case (A), we can see the cancellation between the terms linear to $G_{tt}$ and $G_{ct}$ in $(Y^u_{\pm})_{st}$ when $\tan \beta$ is around 10. Otherwise, the branching ratio deviates significantly from the experimental result unless $m_{H_+}$ is heavier than 500 GeV. The solid gray line draws the prediction of the $G_{ct} = -10^{-3}$ case with $\tan \beta = 50$, which is preferred by the excesses in $R(D^{(*)})$ (see the discussions in the next subsection). The lower bound on the charged Higgs mass reaches $\sim 1$ TeV in this large $\tan \beta$ case. This bound is very strong compared to the one in the type-II 2HDM: $m_{H_{\pm}} > 480$ GeV~\cite{Misiak:2015xwa}. This is because there is a $\tan^2 \beta$ enhancement in Eq.~(\ref{c7}). Note that we have ignored the corrections from the light quark masses, so we need to improve the accuracy related to the light quark masses if a light charged Higgs is observed in future experiments.

In case (B), $\tan \beta$ is fixed to be $\tan \beta = 3$ (blue line), $5$ (red line), and $7$ (green line) in the right panel of Fig.~\ref{fig;BtoXsgamma}. The behavior is totally different from that in case (A) because of the negative sign of $V_{ts}$ in $(Y^u_{\pm})_{st}$. In this case, $(G^u_R)_{ct}$ is negative, so both terms in Eq.~(\ref{eq;Yst}) have the same sign unless $\tan \beta$ is small. As shown in Fig.~\ref{fig;BtoXsgamma}, $\tan \beta \simeq 5$ realizes the cancellation in $(Y^u_{\pm})_{st} $ and allows the charged Higgs mass to be as light as $\sim 300$ GeV, which is distinctly different from the usual type-II 2HDM.

In the next subsection, we discuss $B \to D^{(*)} \tau \nu$, where the deviations from the SM predictions have been reported in the experiments, as shown in Table~\ref{table3}. The excess may require a light charged Higgs boson. Therefore, we study the semileptonic $B$ decay, $B \to D^{(*)} \tau \nu$, in each case, and we discuss our predictions of the observables in the processes.

\begin{table}[t]
  \begin{center}
    \begin{tabular}{|c|c|c|}
      \hline
      Experiment & ~~$R(D)$~~ & ~~$R(D^*)$~~     \\ \hline  \hline
      Belle   & $0.375 \pm 0.064 \pm 0.026 $~\cite{Huschle:2015rga}          & $0.302 \pm 0.03 \pm 0.011 $~\cite{Abdesselam:2016cgx} \\  
      $BABAR$ & $0.440 \pm 0.058 \pm 0.042 $~\cite{Lees:2012xj,Lees:2013uzd} & $0.332 \pm 0.024 \pm 0.018 $~\cite{Lees:2012xj,Lees:2013uzd} \\
      LHCb    &                                                              & $0.336 \pm 0.027 \pm 0.030 $~\cite{Aaij:2015yra} \\
      HFAG    & $0.397 \pm 0.040 \pm 0.028 $~\cite{Amhis:2014hma}            & $0.316 \pm 0.016 \pm 0.010 $~\cite{Amhis:2014hma} \\ \hline
      SM prediction & $0.300 \pm 0.008 $~\cite{RD,RD-2,RD-3,Na:2015kha}      & $0.252 \pm 0.003 $~\cite{RDstar} \\  \hline
    \end{tabular}
  \end{center}
  \caption{Summary of the experimental results in $B\to D^{(*)}\tau\nu$ decays.}
  \label{table3}
\end{table}

\subsection{ $R(D)$ and $R(D^{*})$ }
Next, we investigate the semileptonic decays, $B \to D^{(*)} \tau \nu$, where the deviations from the SM predictions have been reported in the observables concerned with the lepton flavor universality. The interesting measurements are the ratios of the branching ratios for $B\to D^{(\ast)}\tau \nu$ to $B\to D^{(\ast)}l \nu$ ($l=e,\mu$),
\begin{equation}
  R(D^{(*)}) = \frac{\textrm{Br}(B \to D^{(*)} \tau \nu)}{\textrm{Br}(B \to D^{(*)} l \nu)}.
\end{equation}
In the SM, $R(D)=0.300\pm 0.008$~\cite{RD,RD-2,RD-3,Na:2015kha} and $R(D^*)=0.252\pm 0.003$~\cite{RDstar}. The experimental results and the SM predictions are summarized in Table~\ref{table3}, where we find that the discrepancies between them are more than  $2\sigma$.

The semileptonic $b\rightarrow c$ decays are given by the following operators:
\begin{equation}
  {\cal H}^{B-\tau}_\textrm{eff}= C^{cb}_\textrm{SM}(\overline{c_L} \gamma_\mu b_L)(\overline{\tau_L}\gamma^\mu \nu_L) +C_R^{cb}(\ov{c_L}b_R) (\ov{\tau_R} \nu_L)+C_L^{cb}(\ov{c_R}b_L )(\ov{\tau_R} \nu_L),
\end{equation}  
where $C^{cb}_\textrm{SM}$ is the Wilson coefficient in the SM and $C^{cb}_{R,L}$ are generated by the charged Higgs exchange  in our model. In Ref.~\cite{Crivellin:2012ye}, the following simplified expressions for $R(D^{(*)})$ were proposed:
\begin{eqnarray}
  R(D)&=& R_\textrm{SM} \left (1+ 1.5 ~\textrm{Re} \left ( \frac{C^{cb}_{R}+C^{cb}_{L}}{C^{cb}_{\rm SM}} \right ) + \left|  \frac{C^{cb}_{R}+C^{cb}_{L}}{C^{cb}_{\rm SM}} \right |^2 \right ), \\
  R(D^*)&=& R^*_\textrm{SM} \left (1+ 0.12 ~\textrm{Re} \left ( \frac{C^{cb}_{R}-C^{cb}_{L}}{C^{cb}_{\rm SM}} \right ) + 0.05 \left | \frac{C^{cb}_{R}-C^{cb}_{L}}{C^{cb}_{\rm SM}} \right |^2 \right ),
\end{eqnarray}
where each Wilson coefficient is at the $B$ meson scale~\cite{Datta:2012qk}. Here, $R^{(*)}_\textrm{SM}$ are the SM predictions. 

\begin{figure}[!t]
  \begin{center}
    {\epsfig{figure=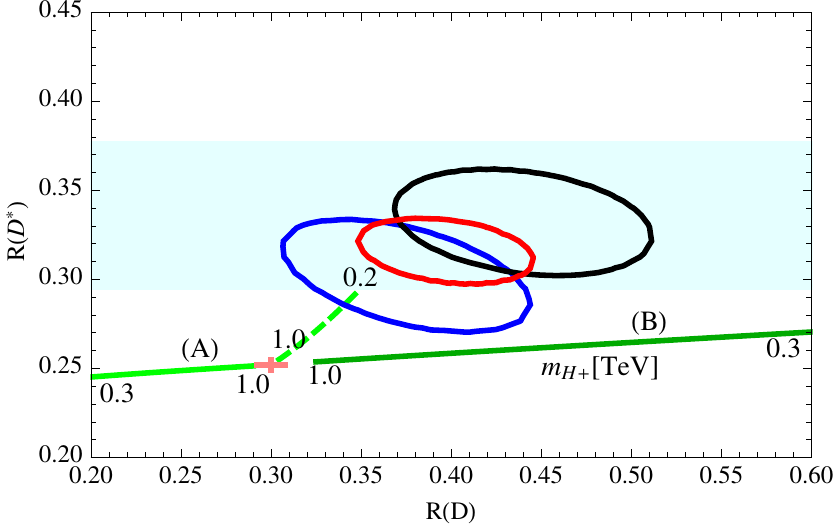,width=0.7\textwidth}} 
    \caption{$R(D)$ vs $R(D^*)$ in cases (A) (light green line) and (B) (dark green line). $\tan \beta$ is fixed at $\tan \beta = 10$ in case (A) and $\tan \beta = 5$ in case (B). The charged Higgs mass is within $300 \, {\rm GeV} \leq m_{H_{\pm}} \leq 1.0 \, {\rm TeV}$ on each line. The dashed line corresponds to the case with $(G^u_R)_{ct}=10^{-3}$, $\tan \beta=50$, and $200 \, {\rm GeV} \leq m_{H_{\pm}} \leq 1.0 \, {\rm TeV}$. Each ellipse describes the $1\sigma$ results for the Belle (blue)~\cite{Huschle:2015rga,Abdesselam:2016cgx}, $BABAR$ (black)~\cite{Lees:2012xj,Lees:2013uzd}, and LHCb (cyan band)~\cite{Aaij:2015yra} experiments and for HFAG (red)~\cite{Amhis:2014hma}, respectively. The pink lines correspond the SM predictions within $1\sigma$~\cite{Na:2015kha,RDstar}.}
    \label{fig;RDandRDs}
  \end{center}
\end{figure}

Integrating out the charged Higgs in our model, we obtain the relevant Wilson coefficients as
\begin{eqnarray}
  C^{cb}_\textrm{SM} &=& 2V_{cb} /v^2, \\
  \label{eq:CL}%
  \frac{C^{cb}_L}{C^{cb}_\textrm{SM}} &=& \frac{m_cm_{\tau}}{m^2_{H_{\pm}}} \tan^2 \beta - \sum_k \frac{V_{kb}}{V_{cb}} \frac{m_k^u m_{\tau} (G^u_R)^*_{kc} }{m^2_{H_{\pm}} \cos^2 \beta}, \\
  \label{eq:CR}%
  \frac{C^{cb}_R}{C^{cb}_\textrm{SM}} &=& - \frac{m_bm_{\tau}}{m^2_{H_{\pm}}} \tan^2 \beta. 
\end{eqnarray}
Figure~\ref{fig;RDandRDs} shows our predictions of $R(D^{(*)})$ in cases (A) and (B). We fix $\tan \beta = 10$ in case (A) and $\tan \beta = 5$ in case (B), respectively, in order to satisfy $B \to X_s  \gamma$ constraint even if the charged Higgs boson mass is in the light region. Each ellipse describes the $1\sigma$ results for the Belle (blue)~\cite{Huschle:2015rga,Abdesselam:2016cgx}, $BABAR$ (black)~\cite{Lees:2012xj,Lees:2013uzd}, and LHCb (cyan band)~\cite{Aaij:2015yra} experiments and for HFAG (red)~\cite{Amhis:2014hma}, respectively. The pink lines correspond to the SM predictions within $1\sigma$~\cite{Na:2015kha,RDstar}.

The light green line corresponds to the prediction of case (A). The charged Higgs mass varies between $300 \, {\rm GeV} \leq m_{H_{\pm}} \leq 1.0 \, {\rm TeV}$ from left to right on the line. When $(G^u_R)_{tc}$ is negative, $C^{cb}_L$ also becomes negative, unless the magnitude of $(G^u_R)_{tc}$ is quite small. Then we find that both $R(D)$ and $R(D^*)$ tend to be smaller than the SM predictions.

The dark green line depicts the prediction of case (B). The charged Higgs mass varies from right to left on the line between $300 \, {\rm GeV} \leq m_{H_{\pm}} \leq 1.0 \, {\rm TeV}$. In contrast to case (A), we see that $R(D)$ in particular can be enhanced significantly compared to the SM prediction. On the other hand, the enhancement of $R(D^*)$ is rather small compared to the experimental results.

The dashed line in Fig.~\ref{fig;RDandRDs} describes the prediction when $(G^u_R)_{ct}=-10^{-3}$ and $\tan \beta=50$ are satisfied. The charged Higgs mass is in the range of $200 \, {\rm GeV} \leq m_{H_{\pm}} \leq 1.0 \, {\rm TeV}$. In this case, large $\tan \beta$ induces an enhancement in $C^{cb}_R$, as well as in $C^{cb}_L$. Both coefficients are comparable, and this parameter choice can enhance both $R(D)$ and $R(D^{*})$. The experimental results, however, require an ${\cal O}(100)$ GeV charged Higgs mass, and the constraint from $B \to X_s  \gamma$ does not allow such a light charged Higgs scenario (see the solid gray line in Fig.~\ref{fig;BtoXsgamma} for the predictions of $B \to X_s  \gamma$ branching ratio).

Recently, it was proposed that the lifetime of $B_c$ can severely constrain the explanation of $B \to D^* \tau \nu$~\cite{Alonso:2016oyd}, which heavily relies on the lattice calculation of the $B_c$ meson decay constant. It is shown that accommodation with $R(D^{*})$ using charged Higgs fields may be in strong tension with the observables in the leptonic decay $B_c\to \tau \nu$~\cite{Alonso:2016oyd}. However, our model can evade the strong bound from $B_c\to \tau \nu$ because it predicts small enhancement in $B\to D^{*}\tau \nu$. On the other hand, our model would be excluded if these excesses in the semileptonic decays $B\to D^{(*)}\tau \nu$ are confirmed to be the signals of new physics in the future.

\subsection{Exotic top decay}
In our model, there are FCNCs involving top and charm quarks, as shown in Eq.~(\ref{GuR}). The mixing parameter $(G_R^u)_{tc}$ is estimated to be about ${\cal O}(0.01)$. However, the FCNCs of the neutral scalars may have enhancements from large $\tan \beta$ and top quark masses, so the flavor-violating top decay, $t\to c h$, can get significantly  large. The effective Lagrangian for  the top decay $t\to c h$  is given by
\beq
\frac{m_t}{v} \tan \beta (G^u_R)_{tc} \left \{ \sin (\alpha-\beta) h \, + \cos (\alpha-\beta) H -iA  \right \} \, \overline{t_L} c_R + {\rm H.c.}
\eeq
The EWPOs require a SM limit where $\cos (\alpha-\beta)$ is close to unity so that the flavor-violating coupling of $h$, whose mass is 125 GeV, vanishes in this limit.

The exotic top decay, $t \to h \, c$, has been investigated at the LHC experiments~\cite{Aad:2015pja,Khachatryan:2016atv}. The upper bound on the Yukawa couplings between top and charm quarks is about $0.1$. Then, our model is still safe for the constraint, as far as $ \sin (\alpha-\beta)$ is less than ${\cal O}(0.1)$.

Note that $(G^u_R)_{tu}$ is smaller than $(G^u_R)_{tc}$ following Eq.~(\ref{GuR}), so that the same-sign top signal, $pp \to tt$, is highly suppressed in the current model, unlike the models  considered in Refs.~\cite{Ko1,Ko2,Ko3}.

\begin{table}[ht]
  \begin{center}
    \begin{tabular}{cccccc}
      \hline
      \hline
      Fields & ~~Spin~~ & ~~$\text{SU}(3)_c$~~ & ~~$\text{SU}(2)_L$~~ & ~~$\text{U}(1)_Y$~~ & ~~U(1)$^{\prime}$~~ \\ \hline  
      $Q^{\prime}_R$     & 1/2 & ${\bf 3}$ & ${\bf2}$ & $1/6$  & $1$         \\ 
      $Q^{\prime}_L$     & 1/2 & ${\bf 3}$ & ${\bf2}$ & $1/6$  & $0$         \\  \hline 
      $u^{\prime}_L$     & 1/2 & ${\bf 3}$ & ${\bf1}$ & $2/3$  & $1$         \\ 
      $u^{\prime}_R$     & 1/2 & ${\bf 3}$ & ${\bf1}$ & $2/3$  & $0$         \\ 
      $u^{\prime\prime}_L$ & 1/2 & ${\bf 3}$ & ${\bf1}$ & $2/3$  & $1 + q_{3}$ \\ 
      $u^{\prime\prime}_R$ & 1/2 & ${\bf 3}$ & ${\bf1}$ & $2/3$  & $0$         \\  \hline
      $R_\mu^{\prime}$    & 1/2 & ${\bf 1}$ & ${\bf2}$ & $-1/2$ & $q_e$       \\ 
      $L_\mu^{\prime}$    & 1/2 & ${\bf 1}$ & ${\bf2}$ & $-1/2$ & $0$         \\ 
      $R_\tau^{\prime}$   & 1/2 & ${\bf 1}$ & ${\bf2}$ & $-1/2$ & $q_e$       \\ 
      $L_\tau^{\prime}$   & 1/2 & ${\bf 1}$ & ${\bf2}$ & $-1/2$ & $0$         \\ \hline 
      $\mu^{\prime}_L$   & 1/2 & ${\bf 1}$ & ${\bf1}$ & $-1$   & $q_e-1$      \\ 
      $\mu^{\prime}_R$   & 1/2 & ${\bf 1}$ & ${\bf1}$ & $-1$   & $0$          \\  
      $\tau^{\prime}_L$  & 1/2 & ${\bf 1}$ & ${\bf1}$ & $-1$   & $q_e-1$      \\ 
      $\tau^{\prime}_R$  & 1/2 & ${\bf 1}$ & ${\bf1}$ & $-1$   & $0$          \\  \hline  \hline
      $\Phi_l$         & 0   & ${\bf1}$  &${\bf 1}$ & $0$    & $q_e$        \\ 
      $\Phi_r$         & 0   & ${\bf1}$  &${\bf 1}$ & $0$    & $q_e-1$      \\
      \hline \hline
    \end{tabular}
  \end{center}
  \caption{The extra chiral fermions for the anomaly-free conditions with $(q_1, \,q_2)=(0,\,1)$. The bold entries ``${\bf 3}$" (``${\bf 2}$") show the fundamental representation of SU(3) (SU(2)) and ``${\bf 1}$" shows singlet under SU(3) or SU(2).}
  \label{table4}
\end{table}

%%%%%%%%%%%%%%%%%%%%%%%%%%%%%%%%%%
\section{EXTRA MATTER FIELDS}
\label{sec;DM}
%%%%%%%%%%%%%%%%%%%%%%%%%%%%%%%%%%
In our model, we considered extra U(1)$^{\prime}$ flavor symmetry under which the SM fermions are charged, so we have to introduce extra chiral fermions in order to achieve anomaly-free conditions. There are several possibilities for the extra field contents.

For instance, the additional fermions with the charge assignments given in Table~\ref{table4} lead to anomaly-free U(1)$^{\prime}$ gauge symmetry. Now we give the Yukawa couplings for the extra fermions involving $\Phi$ and $H_i$:
\begin{eqnarray}
  V_{{\rm extra}}&=&y_Q^{\prime} \overline{ Q^{\prime}_R} \Phi Q^{\prime}_L +y_u^{\prime} \overline{ u^{\prime}_L} \Phi u^{\prime}_R+ y^{\prime} \overline{ Q^{\prime}_R} \widetilde H_3 u^{\prime \prime}_L +y^{\prime \prime} \overline{ Q^{\prime}_L} \widetilde H_1 u^{\prime \prime}_R + \dots + {\rm H.c.}
  \label{Yukawa2}%
\end{eqnarray}
In order to provide the mass terms for the extra fermions, we may have to introduce extra complex scalars which are SM singlets and carry $U(1)^{\prime}$ charges. For instance, we can write the mass terms for the extra leptons and quarks as
\begin{eqnarray}
  y^{\prime}_\mu \overline{R_\mu^{\prime}} \Phi_l L_\mu^{\prime} + y^{\prime}_\tau \overline{R_\tau^{\prime}} \Phi_l L_\tau^{\prime} + y^{\prime \prime}_\mu \overline{\mu_L^{\prime}} \Phi_r \mu_R^{\prime} + y^{\prime\prime}_\tau \overline{\tau_L^{\prime}} \Phi_r \tau_R^{\prime} + y^{\prime \prime}_u \overline{u_L^{\prime \prime}} \Phi^\dagger_r u_R^{\prime \prime},
\end{eqnarray}
where $q_e = -q_3$ is assumed.

In this setup, $Q^{\prime}_L$ is not distinguished from $Q^i_L$, so we can give the Yukawa terms, $\overline{Q^{\prime}_L} \widetilde H_1 \Hat u^a_R$, which causes the mass mixing between the extra quarks and the SM quarks. We assume that the extra fermions are heavy, so the mixing effect is not relevant to our analysis. In any case, the extra quarks can decay through the mixing.

Similarly, we can also find the mixing terms in the lepton sector. For instance, $L^{\prime}_{\mu, \tau}$ carries the same charge as $\Hat L^1$, so the Yukawa couplings, $\overline{L^{\prime}_{\mu, \tau}} H_1 e_R$, are allowed. If we forbid the mass mixing terms, we will obtain dark matter candidates: that is, the neutral components of $R^{\prime}_{\mu, \tau}$ and $L^{\prime}_{\mu, \tau}$.

Another important issue is how to obtain tiny neutrino masses and large lepton mixing. The flavor symmetry limits the Majorana mass matrix for right-handed neutrino, $\nu^i_R$, if nontrivial U(1)$^{\prime}$ charges are assigned to $\nu^i_R$. When $\nu^i_R$ is neutral under U(1)$^{\prime}$ and $q_e$ is defined as $q_e=-q_3$, we can write the Yukawa couplings for the neutrino masses as
\beq
y^\nu_{1i} \overline{L^1} \widetilde{ H_1} \nu^i_R + y^\nu_{2i} \overline{L^2} \widetilde{ H_3} \nu^i_R+ y^\nu_{3i} \overline{L^3} \widetilde{ H_3} \nu^i_R+{\rm H.c.}
\eeq
Note that the Majorana masses of $\nu^i_R$ are also allowed by the flavor symmetry. A detailed study of the phenomenology involving the exotic fermions is beyond the scope of this paper and is left for future study.

%%%%%%%%%%%%%%%%%%%%%%%%%%%%%%%%%%
\section{SUMMARY}
\label{sec;summary}
%%%%%%%%%%%%%%%%%%%%%%%%%%%%%%%%%%
In this paper, we propose an extension of the SM with U(1)$^{\prime}$ flavor gauge symmetry, motivated by the fermion mass hierarchy and the LHCb anomaly. In our model, U(1)$^{\prime}$ charges are assigned to the SM fermions, and flavored Higgs doublets are introduced to obtain the observed fermion mass hierarchy. The alignment of the Yukawa couplings is controlled by the U(1)$^{\prime}$ flavor symmetry, and the VEV alignment of the Higgs doublets realizes the realistic mass matrices for the observed fermions.

Moreover, the charge assignment in Table~\ref{table1} can evade the strong bounds from the Drell-Yan process and the lepton flavor-violating $\mu$ and $\tau$ decays. We can also explain the LHCb anomaly in the $B \to K^* l l$, without conflict with the $B_s$-$\overline{B_s}$ mixing which is the strongest bound in our model.

In this setup, relatively large ($t,c$) elements of the Yukawa couplings of the neutral scalar Higgs bosons are predicted. This coupling is also related with the ($b,c$) element of the charged Higgs Yukawa coupling, which can affect $R(D^{(*)})$. Therefore we have also investigated $R(D^{(*)})$, where significant deviations from the SM predictions are reported by the Belle and $BABAR$ collaborations. $R(D)$ is easily enhanced by the ($b,c$) coupling, but $R(D^*)$ cannot be large because of the stringent bound from $B \to X_s \gamma$. In fact, the strong constraint from $B \to X_s \gamma$ is very strict if the charged Higgs mass is less than 1 TeV. If we require a vanishing $(Y^u_{\pm})_{st}$, we can discuss a light charged Higgs mass, but we need a more precise calculation, including a light quark mass contribution. We also discussed the flavor-violating top quark decay, $t\rightarrow c h$. If the sensitivity of the LHC experiment to the ($t,c$)-Yukawa coupling reaches ${\cal O}(0.01)$, we could test our model in this process.

Finally, let us briefly comment on the physics associated with Higgs fields. Recently, the search for heavy scalar particles as well as the 125 GeV Higgs measurement has been well developed in the LHC experiments. Our predictions for both the heavy resonances and the 125 GeV Higgs are similar to the ones for the type-II 2HDM with $Z^{\prime}$~\cite{Ko:2015fxa} and small FCNCs involving top quark~\cite{Arhrib:2015maa}.\footnote{For Higgs physics in the type-II 2HDM, see, for instance, Ref.~\cite{Chang:2013ona}.} In our setup, we simply take the SM limit to avoid a conflict with the EWPOs, so our Higgs physics is almost the same as the one in the SM. Thus, the direct search for the scalar fields will be a good process for testing our model, although $\tan \beta \sim {\cal O}(10)$ is currently allowed as long as $m_A$ is heavier than 300 GeV~\cite{Aaboud:2016cre}.

%%%%%%%%%%%%%%%%%%%%%%%%%%%%%%%%%%%%
\section*{Acknowledgments}
%%%%%%%%%%%%%%%%%%%%%%%%%%%%%%%%%%%%
We are grateful to Seungwon Baek, Takaaki Nomura, and Yeo Woong Yoon for useful comments on the subject presented in this paper. This work is supported in part by National Research Foundation of Korea (NRF) Research Grant No. NRF-2015R1A2A1A05001869 (P.~K.), by the NRF grant funded by the Korea government (MSIP) (No. 2009-0083526) through the Korea Neutrino Research Center at Seoul National University (P.~K.), by the Do-Yak project of NRF under Contract No. NRF-2015R1A2A1A15054533 (C.~Y.), and by the Basic Science Research Program of the National Research Foundation of Korea (NRF) funded by the Ministry of Science, ICT \& Future Planning (Grant No. 2017R1A2B4011946) (C.~Y.). The work of Y.~S. is supported by the Japan Society for the Promotion of Science (JSPS) Research Fellowships for Young Scientists, No. 16J08299.

{\it Note Added.}
Recently, the LHCb Collaboration announced new data on $R_{K^*}=\textrm{Br}(B\to K^* \mu \mu)/\textrm{Br}(B\to K^* e e)$~\cite{LHCbnew}, which imply about $(2.2\mathchar`- 2.5)\sigma$ deviation from the SM prediction. Motivated by the new data, a number of model-(in)dependent analyses have appeared~\cite{Altmannshofer:2017yso,DAmico:2017mtc,Becirevic:2017jtw}. We find that the preferred regions in Fig.~\ref{fig;btosll} are consistent with the allowed region in the model-independent analysis, particularly in the region in the second figure of Fig.~3 in Ref.~\cite{DAmico:2017mtc}.

%%%%%%%%%%%%%%%% References %%%%%%%%%%%%%%%%%%%%

\end{document}